\documentclass{article}

\usepackage[article,notstatapress]{sj}

\usepackage{epsfig}

\usepackage{stata}

\usepackage{natbib}
\usepackage{chapterbib}
\bibpunct{(}{)}{;}{a}{}{,}

\usepackage{graphicx,amsmath,amssymb,epsfig}
\usepackage{amsfonts}
\usepackage{geometry}
\usepackage{setspace}
\usepackage{lscape}
\usepackage{rotating}
\usepackage{url}

\begin{document}

\inserttype[st0001]{article}
\author{Chernozhukov, Fernandez-Val, Han and Kowalski}{%
  Victor Chernozhukov\\MIT\\Cambridge, Massachusetts\\vchern@mit.edu
  \and
  Ivan Fernandez-Val\\Boston University\\Boston, Massachusetts\\ivanf@bu.edu
  \and
  Sukjin Han\\UT Austin\\Austin, Texas\\sukjin.han@austin.utexas.edu
  \and
  Amanda Kowalski\\University of Michigan\\Ann Arbor, Michigan\\aekowals@umich.edu
}
\title[CQIV]{Censored quantile instrumental variable estimation with Stata}
\maketitle

\begin{abstract}
Many applications involve a censored dependent variable and/or an endogenous independent variable.  \citet{CFK15} introduced a censored quantile instrumental variable estimator (CQIV) for use in those applications, which has been applied by \citet{K16}, among others.  In this article, we introduce a Stata command, \texttt{cqiv}, that simplifes application of the CQIV estimator in Stata.  We summarize the CQIV estimator and algorithm, we describe the use of the \texttt{cqiv} command, and we provide empirical examples.

\end{abstract}

\section{Introduction}

\citet{CFK15} introduced a censored quantile instrumental variable (CQIV) estimator.  In this article, we introduce a Stata command, \texttt{cqiv}, that implements the CQIV estimator in Stata. Our goal is to facilitate the use of the \texttt{cqiv} command in a wide set of applications. 

Many applications involve censoring as well as endogeneity.  For example, suppose that we are interested in the price elasticity of medical expenditure, as in \citet{K16}.  Medical expenditure is censored from below at zero, and the price of medical care is endogenous to the level of medical expenditure through the structure of the insurance contract.  Given an instrument for the price of medical care, the CQIV estimator facilitates estimation of the price elasticity of expenditure on medical care in a way that addressess censoring as well as endogeneity.

The CQIV estimator addresses censoring using the censored quantile regression (CQR) approach of \citet{P86}, and it addresses endogeneity using a control function approach.  For computation, the CQIV estimator adapts the \citet{CH02} algorithm for CQR estimation.  An important side feature of the \texttt{cqiv} stata command is that it can also be used in quantile regression applications that do not include censoring or endogeneity.

In section \ref{sec:estimation}, we summarize the theoretical background on the CQIV command, following \citet{CFK15}.  In section \ref{sec:command}, we introduce the use of the CQIV command.  We provide an empirical application with examples that involve estimation of Engel curves, as in \citet{CFK15}.

\section{Censored quantile IV estimation}\label{sec:estimation}

We first describe a model of triangular system for the CQIV regression. Suppose $Y$ is an observed response variable obtained by censoring a continuous latent response $Y^{\ast }$ from below at the level determined by the variable $C$. Let $D$ be the continuous regressor of interest, possibly endogenous, and $W$ be a vector of
covariates, possibly containing $C$, and $Z$
is a vector of (possibly discrete) IVs excluded from
the equation for $Y^{\ast }$.\footnote{We consider a single endogenous regressor $D$ in the model and in the \texttt{cqiv} procedure.} We observe $\{Y_{i},D_{i},W_{i},Z_{i},C_{i}\}_{i =
1}^{n}$, a sample of size $n$ of independent and identically
distributed observations from the random vector $(Y,D,W,Z,C),$ which
obeys
\begin{eqnarray}
Y &=& \max (Y^{\ast },C),  \label{eq:model1} \\
Y^{\ast } &=& Q_{Y^{\ast }}(U \mid D,W,V)=X'\beta_0(U), \label{eq:model2} \\
D &=& Q_{D}(V \mid W,Z), \label{eq:model3}
\end{eqnarray}
where $V$ is a latent unobserved
variable that accounts for the possible endogeneity of $D$, $X = x(D,W,V)$ with $x(D,W,V)$ being a vector of transformations of $(D,W,V)$, $Q_{Y^{\ast }}(u \mid D,W,V)$ is the $u$-quantile of $Y^{*}$ conditional on $(D,W,V)$, $Q_{D}(v \mid W,Z)$ is the $v$-quantile of $D$ conditional on $(W,Z)$, and
\begin{eqnarray*}
U&\sim &U(0,1) \mid D,W,Z,V,C, \\
V&\sim &U(0,1)  \mid W,Z,C.
\end{eqnarray*}
This CQIV regression model nests the uncensored case of the quantile IV (QIV) regression by making $C$ arbitrarily small. As an example for the CQIV model, in the Engel curve application of \citet{CFK15}, $Y$ is the expenditure share in alcohol, bounded from below at $C=0$, $D$ is total expenditure on
nondurables and services, $W$ are household demographic
characteristics, and $Z$ is labor income measured by the earnings of
the head of the household. Total expenditure is likely to be jointly
determined with the budget composition in the household's allocation
of income across consumption goods and leisure. Thus, households
with a high preference to consume ``non-essential'' goods such as
alcohol tend to expend a higher proportion of their incomes, and
therefore they tend to have a higher expenditure. The control variable $V$ in
this case is the marginal propensity to consume, measured by the
household ranking in the conditional distribution of expenditure
given labor income and household characteristics. This propensity
captures unobserved preference variables that affect both the level
and composition of the budget. Under the conditions for a two stage
budgeting decision process (\citet{GO59}), where the household first
divides income between consumption and leisure/labor and then decide
the consumption allocation, some sources of income can provide
plausible exogenous variation with respect to the budget shares. For
example, if preferences are weakly separable in consumption and
leisure/labor, the consumption budget shares do not depend on labor
income given the consumption expenditure (see, e.g., \citet{DM80}). This justifies the use of labor income as an
exclusion restriction. 

A simple version of the model \eqref{eq:model1}--\eqref{eq:model3} is as follows:
\begin{eqnarray}
Y^{\ast } &=& \beta_{00}+ \beta_{01}D+\beta_{02}W+\Phi
^{-1}(\epsilon),\text{ \ \ }\epsilon \sim U(0,1) \label{eq:ex1}
\end{eqnarray}
where $\Phi^{-1}$ denotes the quantile
function of the standard normal distribution, and also assume that $(\Phi ^{-1}(V)$,
$\Phi ^{-1}(\epsilon))$ is jointly normal with correlation $\rho_0$. From the properties of the
multivariate normal distribution, $\Phi ^{-1}(\epsilon)= \rho_0 \Phi
^{-1}(V)+(1 - \rho_0^2)^{1/2} \Phi ^{-1}(U)$, where $U \sim
U(0,1)$. This result yields a specific expression for the conditional quantile function of $Y^{\ast}$:
\begin{eqnarray}
Q_{Y^{\ast }}(U \mid D,W,V)=X'\beta_0(U)=\beta_{00}+\beta_{01}D+\beta_{02}W+\rho_0 \Phi ^{-1}(V)+ (1 - \rho_0^2)^{1/2}\Phi ^{-1}(U),\label{eq:ex2}
\end{eqnarray}
where $V$ enters the equation through $\Phi ^{-1}(V)$.

Given this model, \citet{CFK15} introduce
the estimator for the parameter $\beta_0(u)$ as
\begin{equation}\label{eq:cqiv}
\widehat{\beta }(u)=\arg \min_{\beta \in \mathbb{R}^{\dim(X)}}\frac{1}{n}\sum_{i=1}^{n} 1(%
\widehat{S}_{i}^{\prime }\widehat{\gamma}>\varsigma)\rho
_{u}(Y_{i}-\widehat X_{i}^{\prime }\beta),
\end{equation}%
where $\rho _{u}(z)=(u-1(z<0))z$ is the asymmetric absolute loss
function of
\citet{KB78}, $\widehat{X}_{i}=x(D_{i},W_{i},\widehat{V_{i}})$%
, $\widehat{S}_{i}=s(\widehat{X}_{i},C_i),$ $s(X,C)$ is a vector
of transformations of $(X,C)$, $\varsigma$ is a positive cut-off, and $\widehat{V_{i}}$ is an estimator of $%
V_{i}$ which is described below.

The estimator in (\ref{eq:cqiv}) adapts the algorithm of \citet{CH02} developed for the censored quantile regression (CQR) estimator to a setting where there is possible endogeneity. \ As described in \citet{CFK15}, this algorithm is
based on the following implication of the model:
$$
P(Y \leq X'\beta_0(u) \mid X, C, X'\beta_0(u) > C) = P(Y^* \leq X'\beta_0(u) \mid X,C,X'\beta_0(u) > C) = u,
$$
provided that $P(X'\beta_0(u) > C) > 0.$ In other words, $X'\beta_0(u)$ is the conditional $u$-quantile of the observed outcome for the observations for which $X'\beta_0(u) > C,$ i.e., the conditional $u$-quantile of the latent outcome is above the censoring point. These observations change with the quantile index and may include censored observations. \citet{CFK15} refer to them as the ``quantile-uncensored'' observations. The multiplier $1(\widehat{S}_{i}^{\prime}\widehat{\gamma}>\varsigma)$ is a selector that predicts if observation $i$ is quantile-uncensored. For the conditions on this selector, consult Assumptions 4(a) and 5 in \citet{CFK15}.

\texttt{cqiv} implements the censored quantile instrumental variable (CQIV) estimator which is computed using an iterative procedure where each step takes the form specified in equation (\ref{eq:cqiv}) with a particular choice of $1(\widehat{S}_{i}^{\prime
}\widehat{\gamma}>\varsigma)$. We briefly describe this procedure here and then provide a practical algorithm in the next section. The procedure first selects the set of quantile-uncensored
observations by estimating the
conditional probabilities of censoring using a flexible binary
choice model. Since $\{X'\beta_0(u) > C\} \equiv \{P(Y^* \leq C \mid X,C) < u\}$, quantile-uncensored observations have conditional probability of
censoring lower than the quantile index $u$. The linear
part of the conditional quantile function, $X_i'\beta_0(u)$, is estimated by standard quantile
regression using the sample of quantile-uncensored observations. Then, the procedure updates the set of quantile-uncensored observations by selecting those observations with conditional
quantile estimates that are above their censoring points, $X_i'\widehat\beta(u) > C_i$, and
iterate.

\texttt{cqiv} provides different ways of estimating the control variable $V$, which can be chosen with option \texttt{\underbar{f}irststage(\ststring)}. Note that
if $Q_D(v \mid W,Z)$ is invertible in $v$, the control variable has
several equivalent representations:
\begin{equation}\label{define 3 model}
V= \vartheta_0(D,W,Z)\equiv F_{D}(D \mid W,Z) \equiv
Q_{D}^{-1}(D\mid W,Z)\equiv \int_{0}^{1}1\{Q_{D}(v\mid W,Z)\leq
D\}dv,
\end{equation}%
where $F_{D}(D \mid W,Z)$ is the distribution of $D$ conditional on $(W,Z)$. Different estimators of $V$ can be constructed based on parametric or semiparametric models for $F_{D}(D \mid W,Z)$ and  $Q_{D}(V\mid W,Z)$. Let $R=r(W,Z)$ with $r(W,Z)$ being a vector of collecting transformations of $(W,Z)$ specified by the researcher. When \texttt{\ststring} is \texttt{quantile}, a quantile regression model is assumed, where
$Q_{D}(v\mid W,Z)=R^{\prime }\pi_0 (v)$ and
\begin{equation*}
V= \int_{0}^{1}1\{R^{\prime }\pi_0 (v)\leq
D\}dv.
\end{equation*}%
The estimator of $V$ then takes the form
\begin{equation}
\widehat{V}=\tau + \int_{\tau}^{1-\tau}1\{R^{\prime }\widehat{\pi }(v)\leq D\}dv,
\label{eq:fe_qr_est}
\end{equation}%
where $\widehat{\pi }(v)$ is the \citet{KB78} quantile
regression estimator which is calculated within \texttt{cqiv} using the built-in \texttt{qreg} command in Stata, and $\tau$ is a small positive trimming constant that avoids estimation of tail quantiles. The integral in \eqref{eq:fe_qr_est} can be approximated
numerically using a finite grid of quantiles.\footnote{The use of the
integral to obtain a generalized inverse is convenient to avoid
monotonicity problems in $v \mapsto R^{\prime }\widehat{\pi }(v)$
due to misspecification or sampling error. \citet{CFG10} developed asymptotic theory for
this estimator.}  Specifically, the fitted values for pre-specified quantile indices (whose number $n_q$ is controlled by option \texttt{nquant(\num)}) are calculated, which then yields
\begin{equation*}
\widehat{V}_i=\frac{1}{n_q}\sum^{n_q}_{j=1} 1\{R^{\prime }_i\widehat{\pi }(v_j)\leq D_i\}.
\end{equation*} For other related quantile regression models that can alternatively be used, see \citet{CFK15}.

When \texttt{\ststring} is \texttt{distribution}, $\vartheta_0$ is estimated using distribution regression. In
this case we consider a semiparametric model for the conditional
distribution of $D$ to construct a control variable
$$
V  = F_D(D \mid W,Z) = \Lambda(R' \pi_0 (D)),
$$
where $\Lambda$ is a probit or logit link function; this can be chosen using option \texttt{ldv1(\ststring)} where \texttt{\ststring} is either \texttt{probit} or \texttt{logit}. The estimator
takes the form
\begin{equation}\label{eq:fe_dr_est}
\widehat V = \Lambda(R' \widehat{\pi}(D)),
\end{equation}
where $\widehat{\pi}(d)$ is the maximum likelihood estimator of
$\pi_0(d)$ at each $d$ (see, e.g., \citet{FP95}, and
\citet{CFM13}).\footnote{\citet{CFM13} developed asymptotic theory for this
estimator.} The expression \eqref{eq:fe_dr_est} can be approximated by considering a finite grid of evenly-spaced thresholds for the conditional distribution function of $D$, where the number of thresholds $n_t$ is controlled by option \texttt{nthresh(\num)}. Concretely, for threshold $d_j$ with $j=1,...,n_t$,
\begin{equation*}
\widehat V_i = \Lambda(R_i ' \widehat{\pi}(d_j)),\quad \textrm{for $i $'s s.t. $d_{j-1} \leq D_i < d_j$ with $d_0=-\infty$ and $d_{n_t}=\infty$,}
\end{equation*}
where $\widehat{\pi}(d_j)$ is probit or logit estimate with $\tilde{D}_i (d_j)=1\{D_i \leq d_j\}$ as a dependent variable and $R_i$ as regressors.

Lastly, when \texttt{\ststring} is \texttt{ols}, a linear regression model $D=R^{\prime }\pi_0+V$ is assumed and $\widehat V$ is a transformation of the OLS residual:
\begin{equation}
\widehat{V}_{i} =\Phi( (D_{i} - R_{i}^{\prime}\widehat{\pi})/\widehat{\sigma}),\label{eq:ols}
\end{equation}
where $\Phi$ is the standard normal distribution, $\widehat{\pi}$ is the OLS estimator of $\pi_0$, and $\widehat{\sigma}$ is the estimator of the error standard deviation. In estimation of \eqref{eq:cqiv} using \texttt{cqiv}, we assume that the control function $\widehat V$ enters the equation through $\Phi ^{-1}(\widehat V)$. This is motivated by the example \eqref{eq:ex1}--\eqref{eq:ex2}.


\subsection{CQIV algorithm}

The algorithm recommended in \citet{CFK15} to obtain CQIV estimates is similar to Chernozhukov
and Hong (2002), but it additionally has an initial step to estimate the control
variable $V$. This step is numbered as 0 to facilitate comparison
with the \citet{CH02} 3-Step CQR algorithm. \

For each desired quantile $u$, perform the following steps:
\begin{enumerate} 
\item [0.]Obtain $\widehat{V}_i = \widehat{\vartheta}(D_i,W_i,Z_i)$
from (\ref{eq:fe_qr_est}), (\ref{eq:fe_dr_est}) or (\ref{eq:ols}) and construct $\widehat{X}_i=x(D_i,W_i, \widehat{V}_i)$.
\item Select a set of quantile-uncensored observations $J_{0}=\{i:\Lambda(\widehat{S}_{i}^{\prime }\widehat{\delta
})>1-u+k_0\}$,
where  $\Lambda$ is a known link function,  $\widehat S_i = s(\widehat X_i, C_i)$, $s$ is a vector of collecting transformations specified by the researcher, $k_0$ is a cut-off such that $0 < k_0 < u,$ and $\widehat{\delta
} = \arg \max_{\delta \in \mathbb{R}^{\dim(S)}} \sum_{i=1}^n \{1(Y_i > C_i) \log \Lambda(\widehat{S}_{i}^{\prime } \delta) + 1(Y_i = C_i) \log [1 - \Lambda(\widehat{S}_{i}^{\prime } \delta)] \}.$
\item Obtain the 2-step CQIV coefficient estimates:  $
\widehat{\beta }^0(u) = \arg \min_{\beta \in \mathbb{R}^{\dim(X)}} \sum_{i \in J_{0}} \rho _{u}(Y_{i}-%
\widehat{X}_{i}^{\prime }\beta),$ and update the set of quantile-uncensored observations, $
J_{1}=\{i:\widehat{X}_{i}^{\prime }\widehat{\beta }^0(u)>C_{i} +
\varsigma_{1}\}.$
\item Obtain the 3-step CQIV coefficient estimates $%
\widehat{\beta }^1(u)$, solving the same minimization program as in step 2 with $J_{0}$ replaced by $J_{1}$.\footnote{As an optional fourth step, one can update the set of quantile-uncensored observations $J_{2}$ replacing $\widehat{\beta }^0(u)$ by $\widehat{\beta }^1(u)$ in the expression for $J_1$ in step 2, and iterate this and the previous step a bounded number of times. This optional step is not incorporated in \texttt{cqiv} command, as \citet{CFK15} find little gain of iterating in terms of bias, root mean square error, and value of Powell objective function in their simulation exercise.}
\end{enumerate}

\noindent \textbf{Remark 1} (Step 1). To predict the
quantile-uncensored observations,
 a probit, logit, or any other model that
fits the data well can be used. \texttt{cqiv} provides option \texttt{ldv2(\ststring)} where \texttt{\ststring} can be either \texttt{probit} or \texttt{logit}. Note that the model does not need
to be correctly specified; it suffices that it selects a nontrivial
subset of observations with $X_i'\beta_0(u)>C_i$. To choose the
value of $k_0$, it is advisable that a constant fraction of
observations satisfying $\Lambda(\widehat{S}_{i}^{\prime
}\widehat{\delta })>1-u$ are excluded from $J_{0}$ for each
quantile. To do so, one needs to set $k_0$ as the
$q_{0}$th quantile of $\Lambda(\widehat{S}_{i}^{\prime }\widehat{%
\delta })$ conditional on $\Lambda(\widehat{S}_{i}^{\prime
}\widehat{\delta})>1-u$, where $q_{0}$ is a percentage (10\% worked
well in our simulation with little sensitivity to values between 5 and 15\%). The value for $q_0$ can be chosen with option \texttt{drop1(\num)}.

\medskip

\noindent \textbf{Remark 2} (Step 2). To choose the cut-off
$\varsigma_{1}$, it is advisable that a constant
fraction of observations satisfying $\widehat{X}_{i}^{\prime }\widehat{%
\beta }^0(u)>C_{i}$ are excluded from $J_{1}$ for each quantile. \
To do so,
one needs to set $\varsigma_{1}$ to be the $q_{1}$th quantile of $\widehat{X}%
_{i}^{\prime }\widehat{\beta }^0(u) - C_i$ conditional on $\widehat{X}_{i}^{\prime }%
\widehat{\beta }^0(u)>C_{i}$, where $q_{1}$ is a percentage less
than $q_{0}$
(3\% worked well in our simulation with little sensitivity to values between 1 and 5\%). The value for $q_1$ can be chosen with option \texttt{drop2(\num)}.\footnote{In practice, it is desirable that $%
J_{0}$ $\subset $ $J_{1}$. If this is not the case, \citet{CFK15} recommend
altering $q_{0}$, $q_{1}$, or the specification of the regression
models. At each quantile,
the percentage of observations from the full sample retained in $J_{0}$,
 the percentage of observations from the full sample retained in $J_{1}$,
 and the percentage of observations from $J_{0}$ not retained in $J_{1}$
can be computed as simple robustness diagnostic tests. The estimator $%
\widehat{\beta }^0(u)$ is consistent but will be inefficient
relative to the estimator obtained in the subsequent step because it uses a
smaller conservative subset of the quantile-uncensored observations if $q_0 > q_1$.}

\medskip

\noindent \textbf{Remark 3} (Steps 1 and 2).  In terms of the notation of \eqref{eq:cqiv}, the selector of Step 1 can be
expressed as $1(\widehat{S}_i'\widehat{\gamma} > \varsigma_0)$,
where $\widehat{S}_i'\widehat{\gamma} =
\widehat{S}_i'\widehat{\delta} - \Lambda^{-1}(1-u)$ and $\varsigma_0
= \Lambda^{-1}(1-u+k_0) - \Lambda^{-1}(1-u)$. The selector of Step 2
can also be expressed as $1(\widehat{S}_i'\widehat{\gamma} >
\varsigma_1),$ where $\widehat{S}_i = (\widehat{X}_i', C_i)'$ and
$\widehat{\gamma} = (\widehat{\beta}^{0}(u)',-1)'$.

\subsection{Weighted Bootstrap Algorithm}

\citet{CFK15} recommend obtaining standard errors and confidence intervals through either weighted bootstrap or nonparametric bootstrap procedures. We focus on weighted bootstrap here. To speed up the
computation, a procedure is proposed that uses a one-step CQIV
estimator in each bootstrap repetition.

For $b = 1, \ldots, B$, repeat the following steps:
\begin{enumerate}
\item Draw a set of weights $(e_{1b}, \ldots, e_{nb})$
i.i.d from the standard exponential distribution.
\item Reestimate the control variable in the weighted sample,  $\widehat V_{ib}^{e} = \widehat \vartheta_{b}^e(D_i, W_i,
Z_i)$, and construct $\widehat X_{ib}^{e} = x(D_i, W_i, \widehat
V_{ib}^{e})$.
\item Estimate the weighted quantile regression:
$
\widehat{\beta }_{b}^{e}(u) = \arg \min_{\beta \in \mathbb{R}^{\dim(X)}} \sum_{i \in J_{1b}} e_{ib} \rho _{u}(Y_{i}-%
\beta^{\prime}\widehat{X}_{ib}^{e}),  \label{QR}
$
where $J_{1b} = \{i:\widehat{\beta }(u)^{\prime}\widehat{X}_{ib}^{e}
> C_{i} + \varsigma_{1} \},$ and $\widehat{\beta }(u)$ is a consistent estimator
of $\beta_0(u)$, e.g., the 3-stage CQIV estimator $\widehat
\beta^1(u)$.
\end{enumerate}

\noindent \textbf{Remark 4} (Step 2). The estimate of the control
function  $ \widehat \vartheta_{b}^{e}$ can be obtained by weighted
least squares, weighted quantile regression, or weighted
distribution regression, depending upon which \texttt{\ststring} is chosen among \texttt{ols}, \texttt{quantile}, or \texttt{distribution} in option \texttt{firststage(\ststring)}.

\medskip

\noindent \textbf{Remark 5} (Step 3). A computationally less expensive alternative is to set $J_{1b} = J_{1}$ in all the
repetitions, where $J_1$ is the subset of selected observations in
Step 2 of the CQIV algorithm. This alternative is not considered in the \texttt{cqiv} routine, because while it is computationally faster, it sacrifices accuracy.
\medskip

\noindent \textbf{Remark 6} As discussed in \citet{CFK15}, we focus on weighted bootstrap, partly because it has practical advantages over nonparametric bootstrap to deal with discrete regressors with small cell sizes, since it avoids having singular designs under the bootstrap data generating process. The \texttt{cqiv} procedure allows both weighted and nonparametric bootstraps.

\medskip

\noindent \textbf{Remark 7} For a cluster bootstrap procedure with clustered data, the bootstrap weights are generated after treating the cluster unit as the unit at which observations are assumed to be independent. In this procedure, the same weight is drawn for all the observations within each cluster.

\section{The cqiv command}\label{sec:command}

\subsection{Syntax}

The syntax for \texttt{cqiv} is as follows:

\begin{stsyntax}
cqiv
    \depvar\
    \optvarlist\
    [endogvar = instrument]
    \optif\
    \optin\
    \optweight\
    \optional{,
    \underbar{q}uantiles(\stnumlist)
    \underbar{c}ensorpt(\num)
    \underbar{c}ensorvar(\varname)
    \underbar{t}op
    \underbar{u}ncensored
    \underbar{e}xogenous
    \underbar{f}irststage(\ststring)
    firstvar(\varlist)
    nquant(\num)
    nthresh(\num)
    ldv1(\ststring)
    ldv2(\ststring)
    \underbar{cor}ner
    drop1(\num)
    drop2(\num)
    \underbar{viewl}og
    \underbar{co}nfidence(\ststring)
    \underbar{clu}ster(\ststring)
    \underbar{b}ootreps(\num)
    \underbar{s}etseed(\num)
    \underbar{le}vel(\num)
    \underbar{no}robust
    }
\end{stsyntax}

\subsection{Description}

\texttt{cqiv} conducts CQIV estimation. 
This command can implement both censored and uncensored quantile IV estimation either under exogeneity or endogeneity. 
The estimators proposed by \citet{CFK15} are used if CQIV estimation or QIV without censoring estimation are implemented. The estimator proposed by \citet{CH02} is used if CQR is estimated without endogeneity.
Note that all the variables in the parentheses of the syntax are those involved in the first stage estimation of CQIV and QIV.

\subsection{Option}

\subsubsection*{Model}

\texttt{\underbar{q}uantiles(\stnumlist)} specifies the quantiles at which the model is 
estimated and should contain percentage numbers between 0 and 100. 
Note that this is not the list of quantiles for the first stage estimation with the quantile regression specification. 
\\
\\
\texttt{\underbar{c}ensorpt(\num)} specifies the fixed censoring point of the dependent variable, where the default is 0; inappropriately specified censoring point will generate errors in estimation.
\\
\\
\texttt{censorvar(\varname)} specifies the censoring variable (i.e., the random censoring point) of the dependent variable.
\\
\\
\texttt{\underbar{t}op} sets right censoring of the dependent variable; otherwise, left censoring is assumed as default.
\\
\\
\texttt{\underbar{u}ncensored} selects uncensored quantile IV (QIV) estimation.
\\
\\
\texttt{\underbar{e}xogenous} selects censored quantile regression (CQR) with no endogeneity, which is proposed by \citet{CH02}.
\\
\\
\texttt{\underbar{f}irststage(\ststring)} determines the first stage estimation procedure, where \texttt{\ststring} is either
 \texttt{quantile} for quantile regression (the default), \texttt{distribution} for distribution regression (either probit or logit), or
 \texttt{ols} for OLS estimation. Note that \texttt{firststage(distribution)} can take a considerable amount of time to execute.
 \\
 \\
\texttt{firstvar(\varlist)} specifies the list of variables other than instruments that are included in the first stage estimation; default is all the variables that are included in the second stage estimation.
\\
\\
\texttt{nquant(\num)} determines the number of quantiles used in the first stage estimation when the estimation procedure is \texttt{quantile}; default is 50, that is, total 50 evenly-spaced quantiles from $1/51$ to $50/51$ are chosen in the estimation; it is advisable to choose a value between 20 to 100.
\\
\\
\texttt{nthresh(\num)} determines the number of thresholds used in the first stage estimation when the estimation procedure is \texttt{distribution}; default is 50, that is, total 50 evenly-spaced thresholds (i.e., the sample quantiles of \texttt{\depvar}) are chosen in the estimation;
 it is advisable to choose a value between 20 and the value of the sample size.
\\
\\
\texttt{ldv1(\ststring)} determines the LDV model used in the first stage estimation when the estimation procedure is \texttt{distribution}, where \texttt{\ststring} is either
 \texttt{probit} for probit estimation (the default), or \texttt{logit} for logit estimation.
 \\
 \\
 \texttt{ldv2(\ststring)} determines the LDV model used in the first step of the second stage estimation, where \texttt{\ststring} is either \texttt{probit} (the default), or \texttt{logit}.
 
 \subsubsection*{CQIV estimation}
 \texttt{\underbar{cor}ner} calculates the (average) marginal quantile effects for censored dependent variable when the censoring is due to economic reasons such are corner solutions. Under this option, the reported coefficients are the average corner solution marginal effects if the underlying function is linear in the endogenous variable, i.e., the average of $1\{Q_{Y^{\ast }}(u\mid D,W,V)>C\}\partial_{D}Q_{Y^{\ast }}(u\mid D,W,V)=1\{x(D,W,V)'\beta_0(u)>C\}\partial_{D}x(D,W,V)'\beta_0(u)$ over all observations. If the underlying function is nonlinear in the endogenous variable, average marginal effects must be calculated directly from the coefficients without \texttt{corner} option. For details of the related concepts, see Section 2.1 of \citet{CFK15}. The relevant example can be found in Section 3.5.
\\
\\
\texttt{drop1(\num)} sets the proportion of observations $q_0$ with probabilities of censoring above the quantile index that are dropped in the first step of the second stage (See Remark 1 above for details); default is 10.
\\
\\
\texttt{drop2(\num)} sets the proportion of observations $q_1$ with estimate of the conditional quantile above (below for right censoring) that are dropped in the second step of the second stage (See Remark 2 above for details); default is 3.
\\
\\
\texttt{\underbar{viewl}og} shows the intermediate estimation results; the default is no log.

\subsubsection*{Inference}

\texttt{\underbar{co}nfidence(\ststring)} specifies the type of confidence intervals. With \texttt{\ststring} being
 \texttt{no}, which is the default, no confidence intervals are calculated. With \texttt{\ststring} being \texttt{boot} or \texttt{weightedboot}, either nonparametric bootstrap or weighted bootstrap (respectively) symmetric t-percentile confidence intervals are calculated. The weights of the weighted bootstrap are generated from the standard exponential distribution. Note that \texttt{confidence(boot)} and \texttt{confidence(weightboot)} can take a considerable amount of time to execute.
\\
\\
\texttt{\underbar{clu}ster(\ststring)} implements a cluster bootstrap procedure for clustered data when \texttt{confidence(weightboot)} is selected, with \texttt{\ststring} specifying the variable that defines the group or cluster.
\\
\\
\texttt{\underbar{b}ootreps(\num)} sets the number of repetitions of bootstrap or weighted bootstrap if the \texttt{confidence(boot)} or \texttt{confidence(weightboot)} is selected. The default number of repetitions is 100.
\\
\\
\texttt{\underbar{s}etseed(\num)} sets the initial seed number in repetition of bootstrap or weighted bootstrap; the default is 777.
\\
\\
\texttt{\underbar{le}vel(\num)} sets confidence level, and default is 95.

\subsubsection*{Robust check}

\texttt{\underbar{no}robust} suppresses the robustness diagnostic test results.  No diagnostic test results to suppress when \texttt{uncensored} is employed.

\subsection{Saved results}

\texttt{cqiv} saves the following results in \texttt{e()}:

\begin{stresults}
\stresultsgroup{Scalars} \\
\stcmd{e(obs)} & number of observations
\\
\stcmd{e(censorpt)} & fixed censoring point
\\
\stcmd{e(drop1)} & $q_0$
\\
\stcmd{e(drop2)} & $q_1$
\\
\stcmd{e(bootreps)} & number of bootstrap or weighted bootstrap repetitions
\\
\stcmd{e(level)} & significance level of confidence interval
\\
\stresultsgroup{Macros} \\
\stcmd{e(command)} & name of the command: \texttt{cqiv}
\\
\stcmd{e(regression)} & name of the implemented regression: either \texttt{cqiv}, \texttt{qiv}, or \texttt{cqr}
\\
\stcmd{e(depvar)} & name of dependent variable
\\
\stcmd{e(endogvar)} & name of endogenous regressor
\\
\stcmd{e(instrument)} & name of instrumental variables
\\
\stcmd{e(censorvar)} & name of censoring variable
\\
\stcmd{e(firststage)} & type of first stage estimation
\\
\stcmd{e(confidence)} & type of confidence intervals
\\
\stresultsgroup{Matrices*} \\
\stcmd{e(results)} & matrix containing the estimated coefficients, mean, standard errors, and lower and upper bounds of confidence intervals
\\
\stcmd{e(quantiles)} & row vector containing the quantiles at which CQIV have been estimated
\\
\stcmd{e(robustcheck)} & matrix containing the results for the robustness diagnostic test results; see Table 1 below 
\\
\end{stresults}
\\
\footnotesize{*Note that the entry \texttt{complete} denotes whether all the steps are included in the procedure; 1 when they are, and 0 otherwise. For other entries consult the paper.}

\normalsize

\subsection{Examples}

We illustrate how to use the command by using some examples. For the dataset, we use a household expenditure dataset for alcohol consumption drawn from the British Family Expenditure Survey (FES); see \citet{BCK07} and \citet{CFK15} for detailed description of the data. Using this dataset, we are interested in learning how the share of total expenditure on alcohol (\texttt{alcohol}) is affected by (logarithm of) total expenditure (\texttt{logexp}), controlling for the number of children (\texttt{nkids}). For the endogenous expenditure, we use disposable income, i.e., (logarithm of) gross earnings of the head of the household (\texttt{logwages}), as an excluded instrument. The dataset (\texttt{alcoholengel.dta}) can be downloaded from SSC as follows:

\begin{stlog}
. ssc describe cqiv
{\smallskip}
. net get cqiv
{\smallskip}
. use alcoholengel.dta
{\smallskip}
\end{stlog}


\noindent The first line will show the dataset is accessible via the second line of the command. The second and third lines will then download \texttt{alcoholengel.dta} to the current working directory. Given this dataset, we can generate part of the empirical results of \citet{CFK15}:

\begin{stlog}
. cqiv alcohol logexp2 nkids (logexp = logwages nkids), quantiles(25 50 75)
{\smallskip}
\oom
{\smallskip}
\end{stlog}

\noindent Here, \texttt{logexp2} is the squared (logarithm of) total expenditure. Using \texttt{cqiv} command, the QIV estimation can be implemented with \texttt{uncensored} option:

\begin{stlog}
. cqiv alcohol logexp2 nkids (logexp = logwages nkids), uncensored
{\smallskip}
\oom
{\smallskip}
\end{stlog}

\noindent And the CQR estimation with \texttt{exogenous} option:
\begin{stlog}
. cqiv alcohol logexp logexp2 nkids, exogenous
{\smallskip}
\oom
{\smallskip}
\end{stlog}

\noindent Here are other possible examples of the CQIV estimation with different specifications and options. Outputs are all omitted.

\begin{stlog}
. cqiv alcohol logexp2 (logexp = logwages), quantiles(20 25 70(5)90) firststage(ols)
{\smallskip}
. cqiv alcohol (logexp = logwages), firststage(distribution) ldv1(logit)
{\smallskip}
. cqiv alcohol logexp2 nkids (logexp = logwages), firstvar(nkids)
{\smallskip}
. cqiv alcohol logexp2 nkids (logexp = logwages nkids), confidence(weightboot) bootreps(10)
{\smallskip}
. cqiv alcohol nkids (logexp = logwages nkids), corner
{\smallskip}
\end{stlog}

\noindent In order of appearance, the commands conduct: the estimation using OLS in the first stage; the estimation using distribution regression with logistic distribution; the estimation where \texttt{nkids} is the only variable other than the instrument that is included in the first stage estimation; the estimation with two instruments and calculating the confidence interval using the weighted bootstrap; and the estimation calculating the marginal effects when censoring is due to corner solutions. In this particular example, \texttt{logexp2} cannot be included in the first-stage regression when distribution regression is implemented. This is because \texttt{logexp2} is a monotone transformation of \texttt{logexp}, and thus the distribution estimation yields a perfect fit.

\section{Acknowledgments}

We would like to thank Simon J{\"a}ger, Blaise Melly and Sanna Nivakoski for helpful comments.

\bibliographystyle{sj}
\bibliography{cqiv}

\begin{table*}[t!]
\caption{CQIV Robustness Diagnostic Test Results for CQIV
with OLS Estimate of the Control Variable - Homoskedastic Design}
\label{optstat}
\begin{centering}
\includegraphics[width=1\textwidth, height= 0.8\textheight]{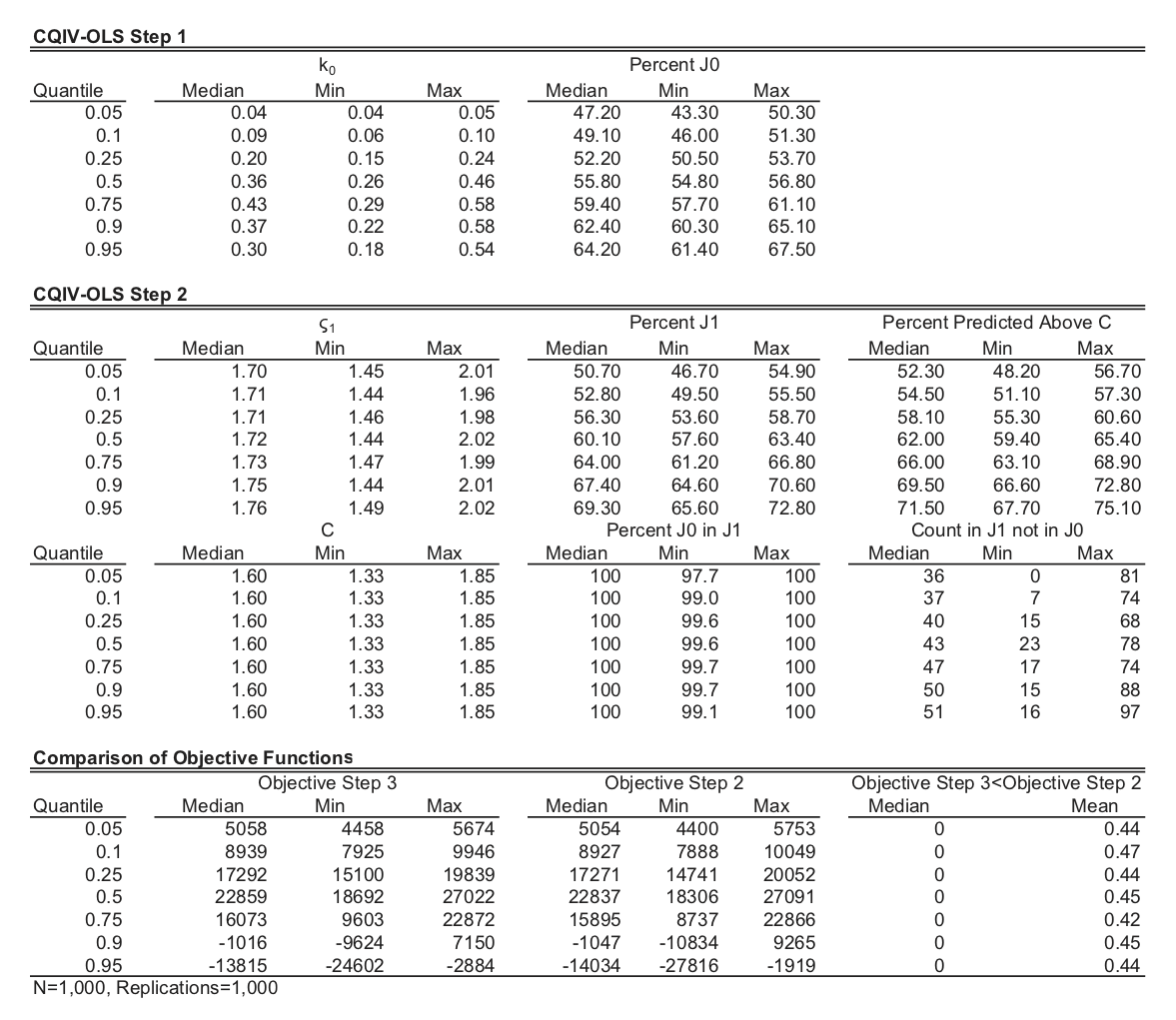}
\end{centering}
\begin{small}In this table, we present the CQIV robustness diagnostic
tests suggested in \citet{CFK15} for the CQIV estimator
with an OLS estimate of the control variable. See Section 2.1 of the present paper for the definitions of $k_0, \varsigma_{1}, J_0, J_1$. In our estimates, we
used a probit model in the first step, and we set $q_0=10$ and
$q_1=3$. In practice, we do not necessarily recommend reporting the
diagnostics in Table \ref{optstat}, but we do recommend examining
them.

In the top section of the table, we present diagnostics computed
after CQIV Step 1. In the second section, we present robustness
test diagnostics computed after CQIV Step 2. In the last section, we report the value of
the Powell objective function obtained after CQIV Step 2 and CQIV
Step 3. See \citet{CFK15} for more discussion.\end{small}

\end{table*}

\end{document}